\documentclass[superscriptaddress,reprint,amsmath,amssymb,aps,prl]{revtex4-2}
\usepackage{lineno,hyperref}
\usepackage[T1]{fontenc}
\usepackage{graphicx,amsmath}
\usepackage{soul,color}
\begin{document}
\title{Breakdown of Charge-Conjugation Symmetry of Disclinations in 2D Crystals}

\author{Ruslan Yamaletdinov}
\email{ruslan.yamaletdinov@epfl.ch}
\affiliation{Institute of Physics, Ecole Polytechnique F\'ed\'erale de Lausanne (EPFL), CH-1015 Lausanne, Switzerland}

\author{Mikhail I. Katsnelson}
\affiliation{Institute for Molecules and Materials, Radboud University, Heijendaalseweg 135, 6525AJ Nijmegen, The Netherlands}
\affiliation{WISE-Wallenberg Inititative Materials Science, Uppsala University, Box 516, SE-751 20 Uppsala, Sweden}

\author{Oleg V. Yazyev}
\email{oleg.yazyev@epfl.ch}
\affiliation{Institute of Physics, Ecole Polytechnique F\'ed\'erale de Lausanne (EPFL), CH-1015 Lausanne, Switzerland}

\date{\today}

\begin{abstract}
Disclinations are elementary topological defects in two-dimensional (2D) crystalline membranes, yet their elastic properties remain largely unexplored. Using atomistic simulations, we address the energetics, morphology, and interactions of disclinations in free-standing graphene, a prototypical 2D crystal. We find that while disclinations with positive topological charges follow an expected behavior, negative disclinations exhibit sublinear energy scaling with charge as well as equilibrium shape that deviates sharply from the conventional saddle ansatz. This breakdown of charge-conjugation symmetry leads to qualitatively distinct interactions: positive disclinations repel, whereas negative disclinations display a robust long-range attraction. These trends are shown to be further amplified by self-adhesion in folded membranes. Our results uncover a fundamentally different energetic landscape for negative curvature defects and provide a basis for understanding the stability, self-folding behavior, and defect-driven morphology of graphene and other 2D membranes.
\end{abstract}

\maketitle

\begin{figure}[b]
\includegraphics[width=1.0\linewidth]{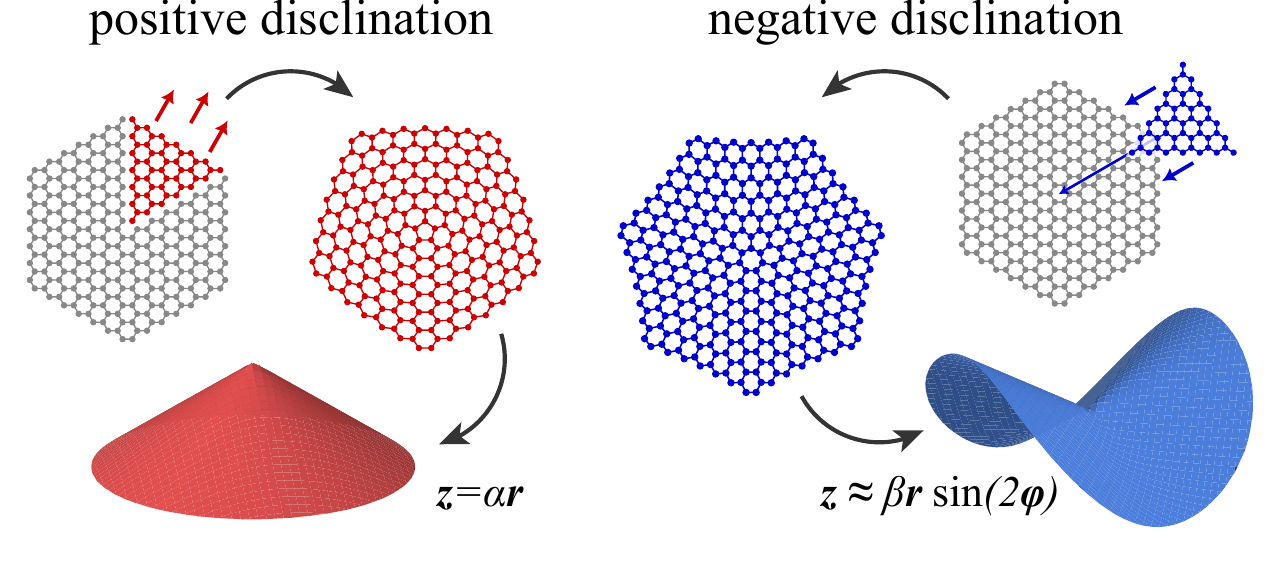}
\caption{Construction of disclinations with topological charges $s=\pm\pi/3$.}
\label{fig:intro}
\end{figure}

{\it Introduction---}A disclination is a topological defect created by inserting or removing a semi-infinite wedge of material in a crystalline lattice~\cite{Volterra1907,Hirth-Lothe}. From a more general geometric point of view disclinations can be considered as sources of curvature~\cite{Katanaev92}. In a two-dimensional (2D) hexagonal network, with graphene being a prototypical example, the simplest disclinations can be constructed by replacing a hexagon by a pentagon or a heptagon, resulting in a positive or negative disclination with topological charges $s=+\pi/3$ and $s=-\pi/3$, respectively (where $s$ measures the angular wedge inserted or removed, see Fig.~\ref{fig:intro})~\cite{Nelson1979,Yazyev2010,Liu2010}, with charge implying a close analogy with electrostatics. Such defects appear in a wide range of $sp^2$-bonded carbon structures~\cite{Georgakilas2015}. An isolated disclination carries a divergent elastic energy and is therefore thermodynamically suppressed~\cite{Hirth-Lothe,Bowick2009}. However, disclinations in finite size membranes or as counterparts of disclination dipoles, that is dislocations, are routinely observed in graphitic materials defining grain boundaries, junctions, and a variety of reconstruction motifs often accompanied by pronounced out-of-plane deformation~\cite{Yazyev2010,Liu2010,Huang11,Kotakoski11,Lehtinen13,Tison2014,Yazyev2014}. The presence of disclinations has nontrivial effects on the electronic structure and is often associated with fractionalization phenomena~\cite{Vozmediano10,Klevtsov14,PhysRevB.88.155127,PhysRevLett.110.046401,PhysRevLett.111.047006,Peterson2021,Liu2021}.

Isolated disclinations can appear in a variety of three-dimensional graphitic systems 
creating non-zero Gaussian curvatures. According to Euler's theorem, the total disclination charge of any closed surface is fixed at $s=4\pi$, which corresponds to at most twelve positive disclinations in an $sp^2$ carbon network, as realized in fullerenes~\cite{Vozmediano10,Brinkmann12,Bille25}, while no analogous topological limit exists for negative disclinations. This distinction becomes particularly important in materials that embed large numbers of negative disclinations. Examples of structures hosting uncompensated or locally uncompensated disclinations include graphene cones, nanotube caps and junctions~\cite{Segawa2016,Gogotsi2015,Shima2018}, crumpled membranes~\cite{Vliegenthart2006,Deng2016,Shima2018}, and periodic negatively curved carbons such as schwarzites~\cite{Segawa2016,MACKAY1991,Shima2018}. In some systems disclinations form spontaneously~\cite{WANG2020144008}; in others, are imposed by the topology of a substrate. For instance, schwarzites might be synthesized via the chemical vapor deposition of a carbon-containing precursor on a zeolite template \cite{Braun2018}. 

In this Letter, we address the elastic and geometric properties of disclinations in free-standing graphene, a prototypical 2D crystal. Using atomistic simulations, we quantify bending energies, equilibrium shapes, and interaction potentials of disclinations of varying topological charge across unfolded, folded, and periodic geometries. We show that negative disclinations violate the conventional energy scaling, adopt distinct equilibrium morphologies, and interact in a manner fundamentally different from their positive counterparts. These results expose a pronounced asymmetry in membrane elasticity and provide a basis for understanding how negative curvature defects govern the stability, morphology, and self-folding or self-adhesion behavior of 2D crystalline materials.

{\it Results---}In flat systems, the presence of a disclination induces a strain field of constant magnitude in the entire system, resulting in the stretching energy of a flat membrane of size $R$ and disclination charge $s$~\cite{Seung1988}:
\begin{equation}
    E_\mathrm{str}=\frac{Y_\mathrm{2D}s^2}{32\pi}R^2,
    \label{eq:E_stretch}
\end{equation}
where $Y_\mathrm{2D}$ is the 2D Young's modulus. 
In the presence of a free dimension, the in-plane stretching energy can be mitigated by out-of-plane deformations. In an unstretchable material, the presence of a disclination introduces a singularity in the Gaussian curvature~\cite{Katanaev92}. In the continuum description, the total bending energy of the membrane~\cite{Seung1988}:
\begin{equation}\label{eq:E_bend}
    E_\mathrm{bend}=\int \left(\frac{1}{2}\kappa H^2 + \kappa_\mathrm{G} K\right) dS,
\end{equation}
where $\kappa$ and $\kappa_\mathrm{G}$ represent the bending and the Gaussian rigidity, $dS$ is the surface element, and $H$ and $K$ are the mean and Gaussian curvatures, respectively. 

In a finite stiffness membrane, a single disclination consists of a stressed core region of radius $R_\mathrm{b}$ surrounded by an outer region that exhibits bending deformation with vanishing in-plane stretching~\cite{Seung1988}. In this outer region, mechanical equilibrium enforces a scale-invariant shape, leading to a logarithmic dependence of bending energy on the system size. Accordingly, the total energy of a membrane of radius $R$ can be written as~\cite{Seung1988,Liu2010}
\begin{equation}
\begin{aligned}
E &= E_{\mathrm{bend}}(r>R_\mathrm{b}) + E_{\mathrm{str}}(R_\mathrm{b}) \\
&= \kappa A(s)\ln\left(\frac{R}{R_\mathrm{b}}\right) + C(s),
\end{aligned}
\label{eq:E}
\end{equation}
where $A(s)$ and $C(s)$ are constants dependent on the disclination charge. The term $C(s)$ represents the energy stored in the disclination core and is approximately the stretching energy accumulated for $0<r<R_\mathrm{b}$. The prefactor $A(s)$ depends sensitively on the resulting equilibrium shape. For positive disclinations, simple cones satisfy the von Kármán equations exactly~\cite{Seung1988}. For negative disclinations, Seung and Nelson proposed the saddle-like form $z(r,\phi)=\beta r\sin(2\phi)$ (Fig.~\ref{fig:intro}) as a variational upper bound on the bending energy~\cite{Seung1988}. This ansatz was shown to yield reasonable estimates in certain geometries~\cite{Deem1996, Park1996,Zhang2014}.

In the case of multiple disclinations, the energetics of flat membranes has been extensively studied~\cite{Richter1987,Romanov2021}. While the total energy depends on the detailed geometry, the interaction term is well established to scale as $s_1 s_2 \ln r_{12}$~\cite{Kosterlitz1974,Halperin1978,Nelson1979}, implying charge-conjugation symmetry and repulsion between like-signed defects in flat space. For suspended membranes, however, existing studies have primarily focused on configurations with zero net disclination charge~\cite{Zhang2014}, and no systematic conclusions have been drawn regarding the interactions between defects of same or opposite signs in the presence of free dimension.

\begin{figure}
\includegraphics[width=0.95\linewidth]{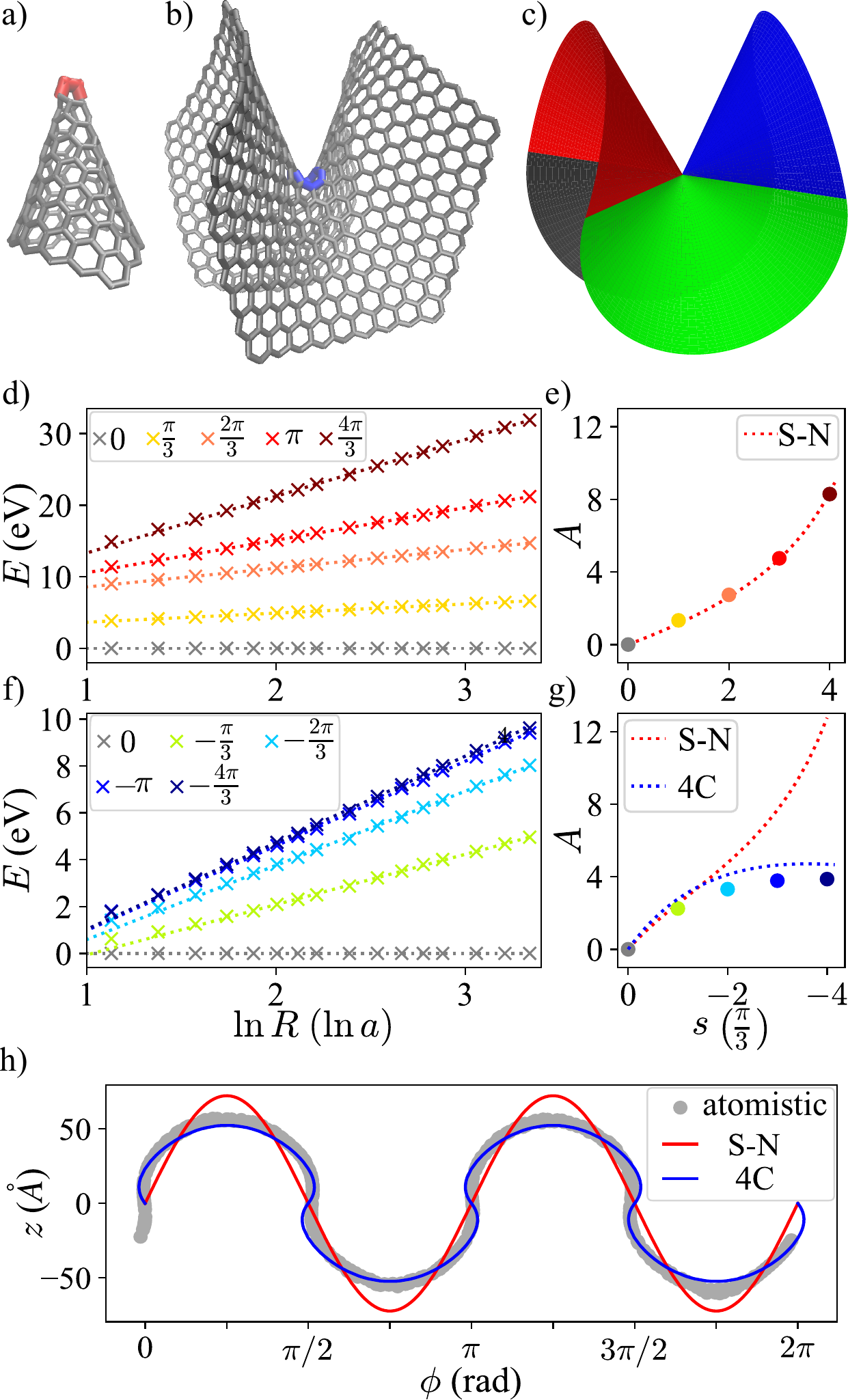}
\caption{
Optimized atomistic geometry containing (a) a positive disclination and (b) a negative disclination with charge $s=\pm 4\pi/3$. Colors highlight the disclination cores. 
(c) Geometric approximation of a negative disclination ($s=-4\pi/3$) by four cones (4C).
(d,f) Energies of circular membranes with disclinations charges (d) $s\ge 0$ and (f) $s\le 0$ plotted as a function of $\ln R$ ($a\approx2.46$~\AA), together with fits of the form $E = \kappa A(s)\ln R + C(s)$.  
(e,g) Prefactor $A(s)$ extracted from the fits  for (e) $s\ge 0$ and (f) $s\le 0$). Bending rigidity $\kappa=0.956$~eV, obtained in line with approach described in Ref.~\cite{Kunihiro2025}.
Red dotted lines denote the predictions of Seung and Nelson (S-N)~\cite{Seung1988}.  
For negative disclinations, blue dotted line shows the estimate obtained from the four-cone approximation (4C).
(h) $z(\phi)$ at constant in-plane radius $r_{xy}=30$~\AA\ for the relaxed atomistic configuration (gray), Seung-Nelson form (red), and the four-cone (4C) model (blue). 
}\label{fig_1}
\end{figure}

Here, we investigate finite-size membranes containing disclinations with disclination charges $|s|\leq\frac{4}{3}\pi$, corresponding to the insertion or removal of up to four $\pi/3$ wedges from an initially flat hexagonal graphene flake (Fig.~\ref{fig_1}(a,b))~\cite{Rozhkov2018}. According to classical force-field atomistic simulations (see Appendix A), and as expected from the continuum theory, both positive and negative disclinations exhibit an energy dependence that scales as $\ln R$ with the system size (Fig.~\ref{fig_1}(d,f); throughout this work, all distances are expressed in units of the lattice constant of pristine graphene, $a\approx2.46$~\AA; the calculated energies are given relative to either hexagonal graphene flakes of matching size, or a pristine graphene layer for models without edges), and the shapes of relaxed configurations remain scale invariant. For positive disclinations ($s>0$), the equilibrium geometry matches a perfect cone, yielding the energy prefactor $A(s)=\frac{s(4\pi-s)}{2(2\pi-s)}$ (see Eq.~\ref{eq:E}) in agreement with the Seung–Nelson prediction~\cite{Seung1988} (Fig.~\ref{fig_1}(e)).

The behavior of negative disclinations is qualitatively different. Their energies grow sublinearly with $|s|$, in contrast to the prediction obtained by applying Eq.~\ref{eq:E_bend} to the Seung–Nelson saddle ansatz $z(r,\phi)=\beta r\sin(2\phi)$ (Fig.~\ref{fig_1}(g)). This discrepancy arises because the saddle surface is not the true solution of the von Kármán equations, but only a variational upper bound on the bending energy~\cite{Seung1988}. This configuration produces excess mean curvature near the four lobes of the shape shown in Fig.~\ref{fig_1}(b,h), leading to a considerable overestimation of the total energy for large negative charges $|s|$.

We propose a more accurate estimate of the negative-disclination geometry can be constructed by distributing the mean curvature as evenly as possible. The simplest such construction consists of four conical sectors sharing a common apex (Fig.~\ref{fig_1}(c), Appendix B), which yields the energy prefactor
\begin{equation}
        A(s)= \frac{2\sin^2\phi-1}{2} (2\pi-s),
\end{equation}
where $\phi$ is determined through
\begin{equation}
        4\sqrt 2 (\pi-\phi)=(2\pi-s) \sin \phi .
\end{equation}
Although this piecewise surface is not a true equilibrium shape---curvature is discontinuous across the the lines joining the conical sectors---it nevertheless provides a significantly better description of both the geometry and the energy at large $|s|$ (Fig.~\ref{fig_1}(g,h)).

\begin{figure}
\includegraphics[width=0.95\linewidth]{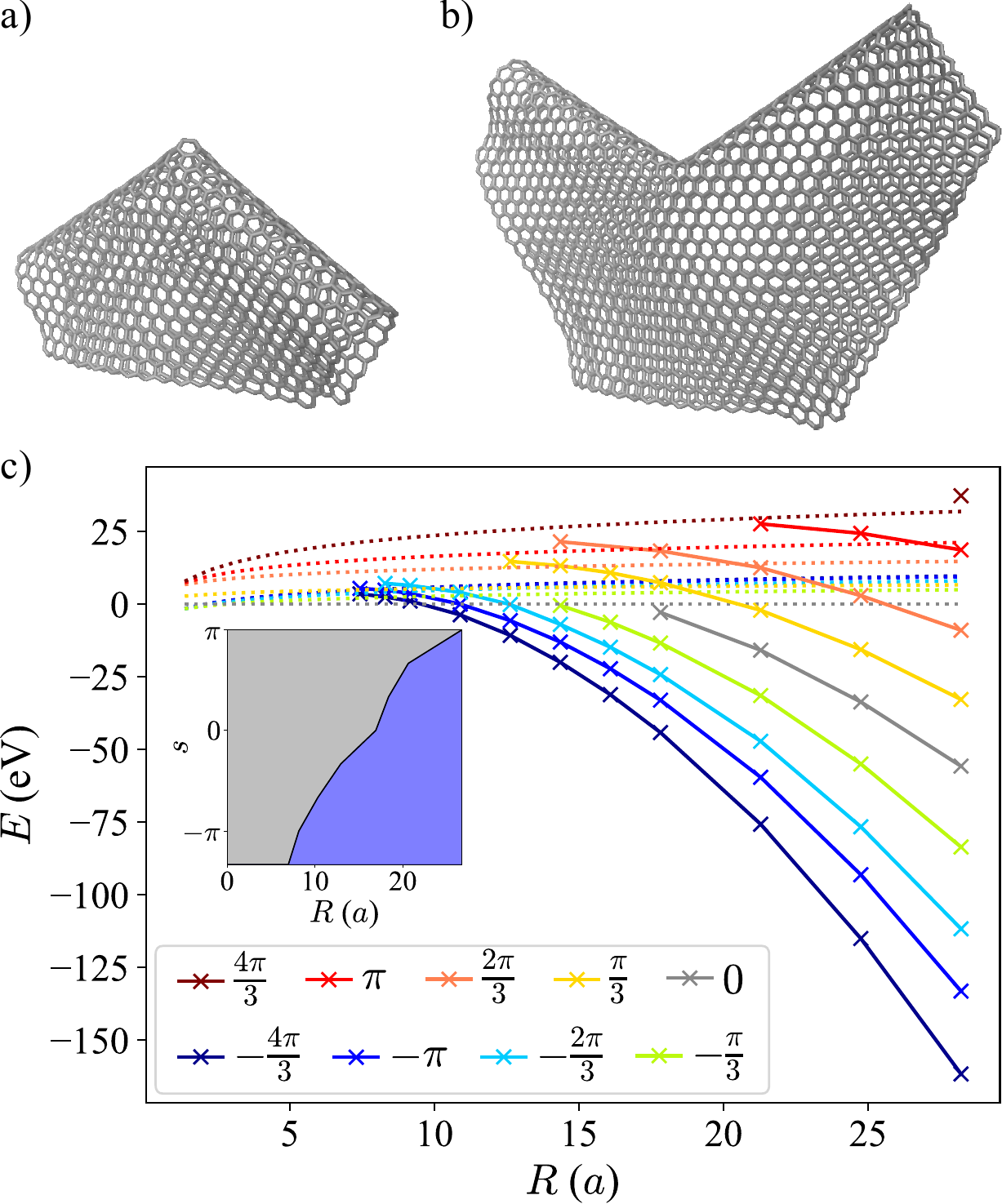}
\caption{
Folded structures containing disclinations with charge  
(a) $s=+2\pi/3$ and (b) $s=-2\pi/3$.  
(c) Energies of folded structures for different disclination charges $s$.  
Dotted lines denote the fits obtained for the corresponding unfolded structures, as in Fig.~\ref{fig_1}(e,g).  
Inset: ground-state structure as a function of system size and disclination charge; gray and blue regions denote unfolded and folded structures, respectively.
}\label{fig_2}
\end{figure}

The behavior of the membrane changes dramatically once folding and van der Waals–driven self-adhesion are allowed. In this regime, large portions of the surface can come into contact with themselves (Fig.~\ref{fig_2}(a,b)), generating a stabilizing adhesion energy of order $\sim \tfrac{1}{2}(2\pi-s)R^2$ per disclination, which rapidly exceeds the energy cost of elastic bending $\sim R + \mathcal{O}(\ln R)+c$~\cite{Meng2013,Yamaletdinov17a}. This excess adhesive energy quickly becomes the dominant factor controlling the overall morphology (Fig.~\ref{fig_2}(c)). For negative disclinations, the associated geometry naturally provides additional surface available for contact, so self-adhesion strongly stabilizes the folded configuration relative to a folded sheet without a defect. In contrast, positive disclinations reduce the area that can participate in adhesion, thereby destabilizing the folded structure compared to the defect-free case.

The asymmetry in the energy scaling for positive and negative disclinations leads to a counterintuitive and important consequence. By analogy with electrostatics, one expects the self-energy to grow superlinearly with the magnitude of the charge $q$, {\it i.e.} $E(2q)>2E(q)$, which ensures that like-signed charges repel. This intuition is indeed correct for positive disclinations, whose energies grow faster than linearly with $s$ in both unfolded and folded geometries. For negative disclinations, however, the sublinear scaling identified above (becoming even decreasing with $|s|$ in folded membranes) implies the opposite trend, suggesting that two negative disclinations attract rather than repel.

\begin{figure}
\includegraphics[width=0.95\linewidth]{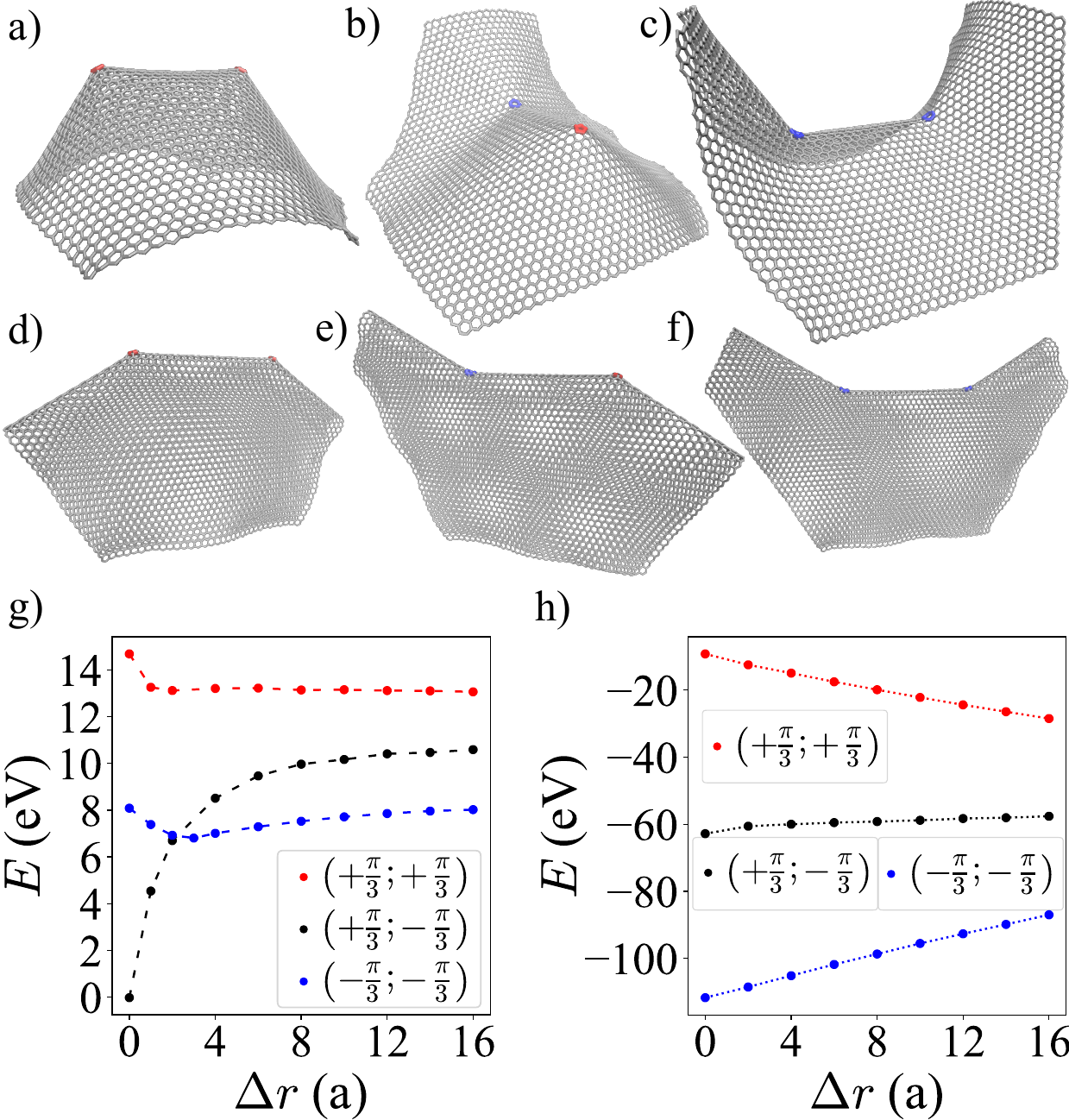}
\caption{
Unfolded and folded configurations containing two disclinations:  
(a,d) two positive disclinations $(+\pi/3, +\pi/3)$,  
(b,e) one positive and one negative disclination $(+\pi/3, -\pi/3)$, and  
(c,f) two negative disclinations $(-\pi/3, -\pi/3)$.  
Total energy as a function of separation $\Delta r$ ($a\approx2.46$\AA) between two disclinations for (g) unfolded  and (h) folded configurations.  
The initial graphene flake has an edge-to-edge diameter of $65 a \approx 16$~nm).
}\label{fig_3}
\end{figure}

To examine this effect directly, we computed the total energies of hexagonal flakes containing two disclinations of charge $s=\pm\pi/3$ placed symmetrically about the center, considering both unfolded and folded geometries (Fig.~\ref{fig_3}(a–f)). For the (++) pair, the unfolded membrane  shows a rapid drop in energy at small $\Delta r$, arising from the strong strain associated with overlapping core regions, followed by an almost flat dependence at larger distances (Fig.~\ref{fig_3}(g), red). For the folded membrane (Fig.~\ref{fig_3}(h), red), the total energy decreases monotonically as the defects are separated.

For the (+–) pair, continuum elasticity predicts strong short-range binding in flat membranes. This behavior is clearly reproduced: the unfolded system shows a steep rise in energy as the defects are separated, which rapidly levels off at larger $\Delta r$ (Fig.~\ref{fig_3}(g), black). In folded membranes (Fig.~\ref{fig_3}(h), black), the short-range core attraction is still present, but the overall dependence becomes nearly linear.

The most notable behavior arises for the (– –) pair. In the unfolded membrane, the energy decreases with decreasing separation over a broad range of $\Delta r$ (Fig.~\ref{fig_3}(g), blue), demonstrating a clear attractive interaction between two negative disclinations. At very small separations, the disclination core regions  overlap, producing a sharp short-range repulsion and an energy minimum at $\Delta r \approx 3a$. The folded membrane exhibits the same qualitative trend (Fig.~\ref{fig_3}(h), blue): bringing two negative disclinations together increases the total adhesion area, further strengthening the attraction.

\begin{figure}
\includegraphics[width=0.95\linewidth]{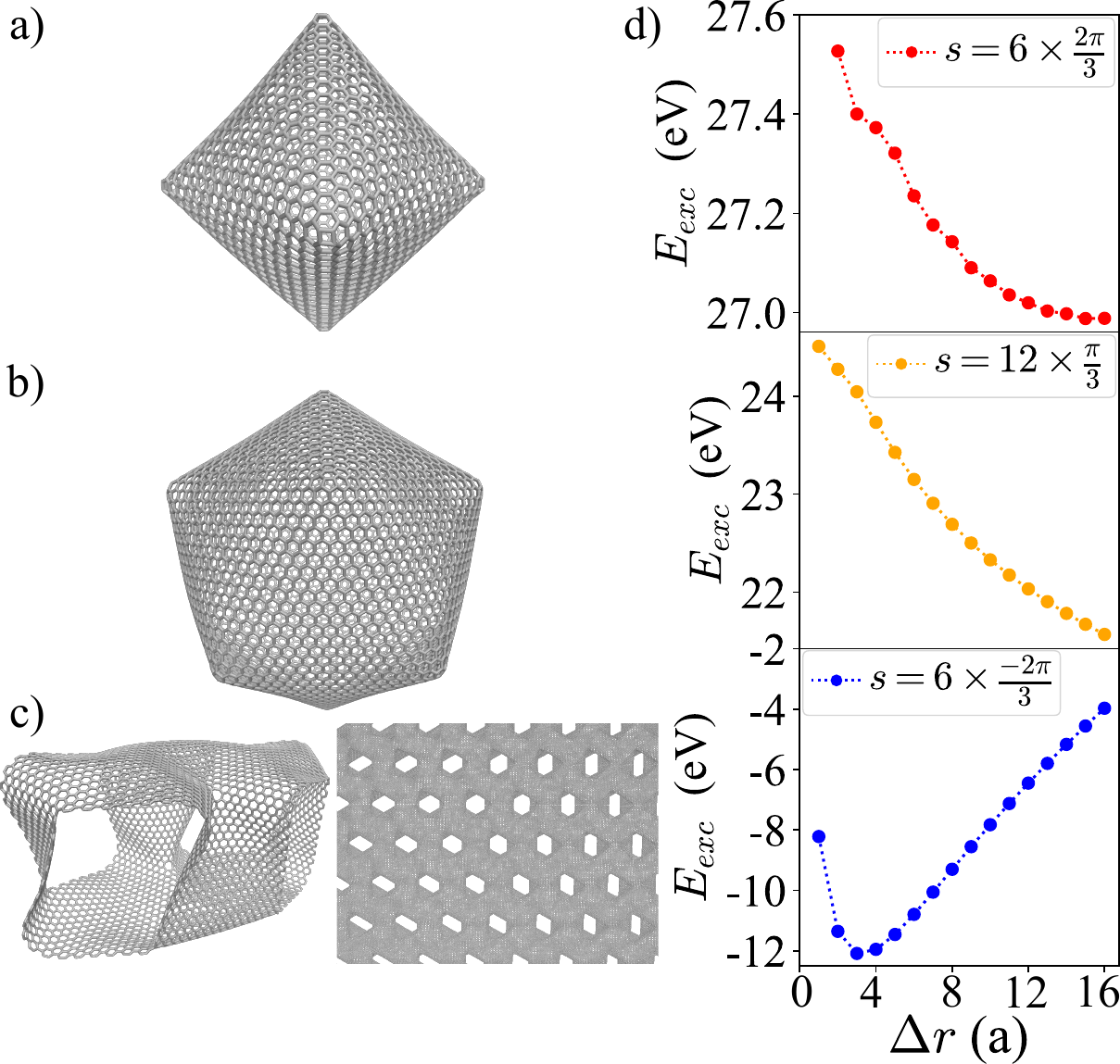}
\caption{
Polyhedral structures formed by (a) six $s=+2\pi/3$ disclinations (octahedron) and (b) twelve $s=+\pi/3$ disclinations (icosahedron), respectively.  
(c) Primitive cell and the top view of a 2D periodic structure obtained for six $s=-2\pi/3$ disclinations, forming a pillared bilayer-graphene geometry with trigonal antiprismatic openings.
(d) Interaction energy $E_\mathrm{exc}$ for structures assembled from equilateral triangular patches of edge length $\Delta r$, containing $n$ identical disclinations of charge $s=\pm n\pi/3$ placed at equivalent positions.
}
\label{fig_4}
\end{figure}

The results discussed above describe the total energies of finite systems. However, when assessing defect–defect interactions, it is more informative to isolate the pair interaction energy. In the geometries considered earlier, edge effects cannot be completely eliminated, and the boundary conditions further complicate the analysis. To mitigate these limitations, we construct closed or periodic structures with no free edges, in which disclinations are placed equidistantly and experience identical local environments. These structures are assembled from equilateral triangular graphene patches of identical size, with each disclination located at a triangle vertex, allowing direct comparison with the single-defect energies of Fig.~\ref{fig_1}.

For positive disclinations, this construction yields regular polyhedra: an octahedron for six disclinations of charge $s=+2\pi/3$ and an icosahedron for twelve disclinations of charge $s=+\pi/3$ (Fig.~\ref{fig_4}(a,b)). For negative disclinations, however, a fully closed structure cannot be assembled using negative defects alone. Nevertheless, for six defects of charge $s=-2\pi/3$ we identify a 2D periodic solution (Fig.~\ref{fig_4}(c)), which forms a pillared, $60^\circ$-twisted bilayer-graphene geometry with trigonal antiprismatic openings. However, we were unable to construct any periodic structure built from equidistant defects of charge $s=-\pi/3$, and whether such a configuration can exist under the same geometric constraints remains for us unclear.

To isolate the interaction energy, we evaluate systems constructed from patches with edge length $\Delta r$ and compute the excess energy relative to isolated disclinations of $\Delta r/2$,
\[
E_\mathrm{exc} = E - \frac{12}{n} A\left(\pm\frac{n\pi}{3}\right) \ln \frac{\Delta r}{2},
\]
where $E$ is the total elastic energy relative to pristine graphene and $A(s)$ is the energy prefactor extracted from Fig.~\ref{fig_1}(f). In this definition we omit the constant term in order to avoid inconsistencies arising from boundary contributions in the single–disclination reference structures.

The resulting interaction energies, plotted in Fig.~\ref{fig_4}(d), clearly reinforce the asymmetry discussed above. Systems composed of positive disclinations ($s>0$) exhibit a monotonically decreasing $E_\mathrm{exc}$ with increasing separation, consistent with long-range repulsion. In contrast, system with negative disclinations ($s<0$) displays a pronounced long-range attraction: $E_\mathrm{exc}$ decreases as $\Delta r$ is reduced, reaching a minimum at $\Delta r \approx 3a$ in full agreement with the results for obtained for finite-size configurations.


{\it Conclusions---}Our results reveal a fundamental asymmetry in the elasticity of free-standing 2D crystalline membranes: positive disclinations follow the classical picture, whereas negative disclinations do not. Atomistic simulations show that the bending energy of negative disclinations grows sublinearly with the absolute value of topological charge $s$, in sharp contrast to the commonly used variational saddle ansatz. This breakdown of charge-conjugation symmetry manifests itself directly in their interactions: positive disclinations repel, while negative disclinations exhibit a long-range attraction with an equilibrium separation of approximately three lattice constants. This asymmetry persists across all geometries examined and is further enhanced  by self-adhesion in folded configurations. Taken together, these findings establish a qualitatively different energetic landscape for negative disclinations and provide a starting point for understanding the thermodynamics and stability of defect-driven curvature in low-dimensional materials. 

These results can be interesting also in a general context of statistical physics of 2D systems where topological defects are supposed to play crucial role \cite{Kosterlitz1974,Halperin1978,Nelson1979}. Graphene plays the role of a fruit fly in this field of science \cite{katsnelson2013,katsnelson2020}, and atomistic simulations tell us that, for example, its melting turns out to be much more complex phenomena than assumed in the basic general models \cite{los2011,los2015}. Our current study demonstrates that disclinations in 2D may show complicated and sometimes counterintuitive phenomena which may stimulate further development of physics of phase transitions in 2D. 

{\it Acknowledgments}--- M.I.K. acknowledges support from the Wallenberg Initiative Materials Science for Sustainability (WISE) funded by the Knut and Alice Wallenberg Foundation (KAW).


\appendix

\section{Appendix A: Simulation details}

To obtain the equilibrium shapes and energies reported in this work, we performed large-scale geometry

optimizations using a two-stage classical force-field protocol. In the first step, to avoid unphysical rebonding during initial relaxation, we employed a CHARMM-like potential that  reproduces the in-plane stiffness and bending rigidity of graphene~\cite{Yamaletdinov2017}. The relaxation was carried out using the OpenMM toolkit~\cite{Eastman2017}. The resulting structures were further relaxed using the AIREBO bond-order force field~\cite{airebo} as implemented in LAMMPS~\cite{Gissinger24}. All LAMMPS calculations were converged with an ultra-tight energy tolerance, with the change in energy between outer iterations is less than $10^{-15}$.

The calculated energies are given relative to either hexagonal graphene flakes of matching size (constructed so that their sectors correspond to those used in constructing the target structure, with identical numbers of edge and bulk atoms), or a pristine graphene layer for models without edges.

\section{Appendix B: Energy of a configuration with a negative disclination composed of joined cone sectors}
\label{sec:cones}

To estimate the energy of structures with negative disclinations at large disclination charge $|s|$, we consider a simple idealized geometry in which the curvature is distributed evenly along an equiradial line. Strictly speaking, such a surface cannot exist: a saddle-like geometry necessarily involves alternating upward- and downward-directed deflections, and the requirement of a continuous surface gradient precludes a perfectly uniform curvature distribution. Nevertheless, this construction provides a useful variational estimate for the energy of structures with large negative disclination charge.

We approximate the surface by four identical circular conical patches whose apices are located at $(0,0,0)$. The cones are aligned along the $x$ and $y$ axes and are displaced along the $z$ direction. Let $\theta$ denote the apex half-angle of each cone. This angle is related to the parameter $\alpha$ introduced in Eq.~(4.24) of Ref.~\cite{Seung1988} via $\alpha = \cot \theta$.

By symmetry, all junctions between neighboring conical patches lie in the $xy$ plane. We further denote by $2\phi$ the angular extent of the arc removed from each cone to enable their assembly into a saddle-like configuration. Conservation of the total circumference of the membrane at a fixed radial distance imposes the constraint
\begin{equation}
    8 (\pi - \phi)\sin\theta = 2\pi - s ,
\end{equation}
where $s$ is the total disclination charge.

A second constraint follows from surface continuity at the junctions between neighboring cones. In particular, adjacent conical patches must meet at right angles along their common boundaries in the $xy$ plane. This geometric condition yields
\begin{equation}
    \sin\theta \, \sin\phi = \frac{1}{\sqrt{2}} .
\end{equation}

Together with the circumference constraint, these relations can be rewritten as
\begin{equation}
    \begin{aligned}
        &4\sqrt 2 (\pi-\phi)=(2\pi-s) \sin \phi\\
        &\alpha^2=2\sin^2\phi-1
    \end{aligned}
\end{equation}

Using the bending energy of a simple conical structure $E_c$ (Eq.~(4.24) of Ref.~\cite{Seung1988}), we can now estimate the total energy of the four-cone configuration
\[
E=4\frac{\pi-\phi}{\pi} E_c=\kappa \frac{\alpha^2}{2} (2\pi-s)\ln R.
\]

\bibliography{biblio}
\end{document}